# Surface optical phonon modes in ternary aligned crystalline InGaN/GaN multi-quantum well nanopillar arrays

S.-E. Wu,[1,2] S. Dhara,[3,4] T.-H. Hsueh,[5] Y.-F. Lai,[1] C.-Y. Wang,[1] and C.-P. Liu[a),1,6]

[1] Department of Materials Science and Engineering, National Cheng Kung University, Tainan 70101, Taiwan

[2] Genesis Photonics Inc., Tainan Science-Based Industrial Park, Tainan 741, Taiwan

[3] Department of Electrical Engineering, Institute of Innovations and Advanced Studies, National Cheng Kung University, Tainan 70101, Taiwan

[4] Materials Science Division, Indira Gandhi Centre for Atomic Research, Kalpakkam 603102, India

[5] Institute of Electro-Optical Science and Engineering, National Cheng Kung University, Tainan 70101, Taiwan

[6] Center for Micro/Nano Science and Technology, National Cheng Kung University, Tainan 70101, Taiwan

## Abstract

The optical properties of focused ion beam-engraved perfectly aligned and spatially controlled multi-quantum-well InGaN/GaN nanopillars were investigated. Crystalline MQW nanopillars with a diameter of 30–95 nm and high aspect ratios (7:1–16:1) showed a maximum of three-fold enhancement in emission intensity per unit active area. A light emitting contour map of Taiwan is drawn with a nanopillar structure to demonstrate the site control of the technique adopted in the present study. Raman scattering studies were used to characterize the newly created surfaces. Unknown peaks in GaN and InGaN nanostructures are identified for surface optical (SO) phonon modes with proper assignments of wave vectors using multiple excitations, and the SO mode for the ternary phase is reported for the first time.

KEYWORDS: surface optical mode, Raman scattering, focused ion beam, GaN, InGaN

a) Corresponding author E-mail: cpliu@mail.ncku.edu.tw

INTRODUCTION

White light emitting diodes (LEDs), which consist of an InGaN blue LED pumping a yellow phosphor, are crucial in solid-state lighting (SSL). Micro/nano-scale InGaN LEDs have been investigated for use in high efficiency SSL, owing to their increased internal quantum efficiencies and increased extraction efficiencies.[1,2] Several studies, showing only orientational control for the nano-LEDs, have proposed the enhancement of either photoluminescence (PL)[3] or electroluminescence (EL)[4,5] in InGaN/GaN multi-quantum well (MQW) 1-D nanostructures. Formation of perfectly aligned nanostructures with good site control is important for optoelectronic applications for nanodevices. Focused ion beam (FIB) technology focuses the ion beam at the nanoscale (~5–10 nm), and has several advantages over other high-energy particle beams used in nanofabrication.[6]

Raman scattering is used to probe the surface optical (SO) phonon mode of nanostructures.[7] In polar crystals, SO phonon modes are observed on the low wavenumber side of the longitudinal optical (LO) phonon mode. For nanocrystals of only a few nanometers, lattice vibration is confined to the surface, giving rise to a wavenumber in between the transverse optical (TO) and LO phonon modes.[8] However, the SO phonon modes cannot be observed optically for a perfect surface, due to momentum conservation constraints. Surface roughness or construction of a grating capable of absorbing the phonon momentum is warranted for perturbing the surface potential and to make the SO phonon observable.[9] To date, there are very few reports available on the Raman study of SO phonons in nanostructured III–V nitride (GaN, InN) systems. In a recent study, Hsiao *et al.* report on a SO phonon at 708.7 cm$^{-1}$ in a single one dimensional (1-D) GaN nanowire (NW) as a broad hump-like feature for 532 nm laser line excitation.[10] In addition, a GaN nanopillar of diameter ~

50 nm is reported to show a SO phonon mode ~716 cm$^{-1}$ with the excitation of 488 nm laser line.[11] Recently, we have reported SO phonon modes of 528 cm$^{-1}$ and 560 cm$^{-1}$ for InN nanostructures with the excitation of a 532 nm laser line.[12] However, there are hardly any reports for the SO phonon mode in the very important ternary phase of InGaN, which is used for many SSL applications.[1,2]

In this study, the fabrication of aligned GaN nanopillars with spatial control of embedded InGaN/GaN MQWs is achieved using FIB milling. We report here the first identification of the SO phonons in the 1-D InGaN/GaN MQW system, with detailed calculations considering their stoichiometric compositions.

**EXPERIMENTAL**

The epitaxial (epi-) film used in this study was an LED structure grown on a *c*-plane sapphire ($\alpha$-Al$_2$O$_3$) by metal-organic chemical-vapor deposition, composed of a 30-nm-thick GaN buffer layer, a 4-µm-thick Si-doped *n*-GaN layer, 10-pair In$_y$Ga$_{1-y}$N/GaN MQWs, and a 560-nm-thick Mg-doped *p*-GaN. The FIB instrument used for patterning was an SMI3050SE FIB-SEM (scanning electron microscope) hybrid system from SII NanoTechnology Company. Wet chemical etching using KOH with 20 wt% dissolved in ethylene glycol (EG) was used as the etchant, with an etching time of less than 3 min, to remove the outer swollen volume of the pillars after FIB milling. FIB assisted dry etching using reactive gas XeF$_2$ was also tried for the formation of specific structures, because it was capable of reducing the process time by minimizing the unwanted ion fluence leading to ion beam-induced damage. A room-temperature cathodoluminescence (CL) measurement was performed on a JEOL JSM-7000F field-emission SEM equipped with a Gatan MonoCL. A micro-Raman spectroscopic study, using 532 and 632.8 nm excitation, was carried out with a Jobin-Yvon LabRam HR fitted with 1800 lines per mm gratings (spectral resolution ~0.5 cm$^{-1}$). All the spectra were taken in the back-scattering configuration, denoted as [$z(x,x)\bar{z}$], with the $z$ direction parallel and the $x$ direction perpendicular to the $c$ axis of the epi-layer sample. A liquid nitrogen cooled charged coupled device (CCD) detector was used for collecting the scattering intensity. The spectra were acquired from the pillars at room temperature using a microscope with a focal spot of 1 μm, which was small enough to probe the pillar-distributed area. High resolution transmission electron microscopy (TEM) analysis was also conducted to analyze the pillar microstructure with a JEOL JEM-2100F, operated at 200 keV.

## RESULTS AND DISCUSSION

### MICROSTRUCTURAL AND LUMINESENCE STUDIES

A 30 keV Ga$^+$ beam with a current of 300 pA was employed to completely remove the epi-layered nitride materials around the pillars. The scanning electron microscopy (SEM) image showed the resulting structures (Fig. 1) milled with a serpentine-scanning beam at a dwell time of 150 μs with an ion fluence of 3.2×10$^{17}$ cm$^{-2}$ and subsequently wet-etched with KOH solution. The pillar base diameter in Fig. 1a is around 95 nm (aspect ratio 7:1). With an extended KOH treatment time, the pillar size can be further reduced to 30 nm (Fig. 1b) with enhanced high-aspect-ratio (16:1).

The CL spectra from the pillars, before and after chemical etching, are shown in Fig. 2a. A typical SEM top view image is shown in Fig. 2b for the nanopillar assembly with a corresponding spectroscopic CL image at the specific wavelength of 418 nm. It should be noted that it was very difficult to observe even a reasonable CL image from nanostructures with a diameter ~ 95 nm. This is primarily due to the large uncompensated surface states in the small sized nanoclusters. Capturing excellent CL images of small sized 1-D nanostructures with diameters below 100 nm was possible due to the surface state compensation of the chemical treatment used in this technique. The intensity of the MQW emission from the wet-etched pillars is enhanced by a factor of 13 compared with that of the as-milled pillars. When compared with the as-grown epi-layer sample, the arrayed pillars have a three-fold enhancement in MQW emission intensity. This value was estimated after being normalized by a much smaller effective quantum well area, estimated from 53 pillars in the field acquired by the CL apparatus at one time. The estimated enhancement in luminescence property is comparable to other reported values for similar structures with random arrays.[3–5] With almost no blue-shift found, it is proposed that light scattering off the sidewalls of the nanopillars boosts the total emission intensity after the swollen volume is efficiently etched out.

We also demonstrate the absolute site-controlling capability, particularly on the nanoscale, by creating a contour map of Taiwan. This is performed with high aspect-ratio pillars distributed in the bitmap-defined area (i.e. the black pixels), as shown in Figs. 3a–c. A light-emitting Taiwan in the corresponding CL image at the specific wavelength of 420 nm can clearly be seen (Fig. 3d), since each pillar preserves the perfection of the quantum well structure,[13] and is able to emit light efficiently once the surface damage layer is cleaned by reactive gas (e.g., $XeF_2$) or wet etching.

**LATTICE VIBRATIONAL PROPERTIES**

Fig. 4 shows the Raman spectra recorded from the nanopillar array before and after KOH solution treatment. The spectra are also compared with that of the pristine un-patterned MQW wafer (Fig. 4a) with a symmetry allowed strong peaks near 568 cm$^{-1}$, assigned to be $E_2$(high) phonon mode of wurtzite GaN,[14] and a weak peak near 559 cm$^{-1}$ corresponding to $E_2$(high) phonon mode of InGaN structure when excited with a 532 nm laser line.[15] Other symmetry allowed mode, $A_1$(LO) phonon is also observed near 734 cm$^{-1}$. The Raman mode for sapphire substrate is also observed around 417 cm$^{-1}$.[16] The spectrum is also collected for the excitation of a 632.8 nm laser line for reference where an additional peak around 640 cm$^{-1}$ is reported as the zone boundary (ZB) phonon for cubic mode GaN.[17] The appearance of the ZB phonon in the case of 632.8 nm excitation may be due to the different focusing conditions and spot selected in the sample where defects present are high, as the cubic phase is reported to be present in the stacking faults for GaN.[18] To clarify it further that defect density is relatively higher close to the unetched epi-GaN layer which do not contribute to the luminescence process. The peak at 734 cm$^{-1}$ (Fig. 4b) can not be observed clearly in the as-milled pillars with swollen blisters, possibly due to the broadened features in the spectra indicating the lack of bonding order in the damaged pillar shell. Interestingly, the spectrum of the KOH treated pillars (acquired from those in Fig. 1a) displays distinct Raman features (Figs. 4c, and 4d). All the Raman modes in the pristine wafer (Fig. 4a) are also observed for the wet etched nanopillar sample with 632.8 nm excitation (Fig. 4c), along with the appearance of peaks around 270, 530-570 and 687 cm$^{-1}$. The peak around 270 cm$^{-1}$ may be assigned as ZB phonon of GaN ( indicated as ZB$_{GaN}$ in Fig. 4c) belonging to the *M*-symmetry point of Brillouin zone in the wurtzite GaN[19] as a possible effect of finite sized crystal. Peaks at 534 and 545 cm$^{-1}$ (inset in Fig. 4c) may be attributed closely to the forbidden (in the backscattering configuration) TO phonon in the $A_1$ and $E_1$ symmetries of GaN and InGaN phases, respectively.[14,15] Appearance of the TO phonons is due to the increased contribution of scattering from other planes of GaN, which are exposed on nanopillar sidewalls. This statement is made in the pretext that the scattering is also contributed from the other planes even if the Raman measurements are performed in backscattering configuration. The peaks around 665–699 cm$^{-1}$ can not be assigned to the known Raman modes of GaN. Similar features can also be found in the spectrum for 532 nm excitation (Fig. 4d). Additional peak at 717 cm$^{-1}$ can primarily be assigned to the $A_1$(LO) phonon mode of strain-relaxed In$_y$Ga$_{1-y}$N embedded in the nanopillar, because for $y$=0.15 (In$_{0.15}$Ga$_{0.85}$N, as in our case) the $A_1$(LO) mode for the strained InGaN is reported to be around 729 cm$^{-1}$,[20,21] while the one for relaxed InGaN should be within the range of 711–717.5 cm$^{-1}$.[20,22,23] However, the broad peak in the range of 665–699 cm$^{-1}$ is again a new feature with respect to the pristine film. The broad peak in the range of 625-775 cm$^{-1}$ (inset in Fig. 4d), fitted for various peaks, is shown in the inset.

Depending on the excitation energy, peaks around 685–705 cm$^{-1}$ are also reported as a specific 'S' mode for various compositions of In$_y$Ga$_{1-y}$N without any physical implications.[15,24,25] In view of these ambiguous assignments of the observed unknown Raman modes around 665–699 cm$^{-1}$ (including the unidentified peak around 687 cm$^{-1}$ in Fig. 4c) we examine the possibility of surface phonon being responsible for the new modes. Owing to the availability of surface roughness or even surface modulation in the 1-D MQW nanopillar (Fig. 5), we anticipate these unidentified peaks may originate from the SO phonon modes. The detailed analysis considering surface modulation is performed later. The wavenumber corresponding to SO phonon mode and intensity mainly depend on the size and the shape of the nanostructured material. The expression for SO phonons for a 1-D nanostructure is given by,[8]

$$\omega^2{}_{SO} = \omega^2{}_{TO}\frac{\varepsilon_0 - \rho_{nx}\varepsilon_m}{\varepsilon_\infty - \rho_{nx}\varepsilon_m} \quad (1)$$

where $\omega_{TO}$ is the wavenumber of the TO phonon mode, $\varepsilon_0$ and $\varepsilon_\infty$ and are the static and high frequency dielectric constant of the material, and $\varepsilon_m$ is the dielectric constant of the medium. $\rho_{nx}$ is given by

$$\rho_{nx} = \frac{K_1(x)I_0(x)}{I_1(x)K_0(x)} \quad (2)$$

where $I_j$, and $K_j$ are the modified Bessel functions and $x=qr$ ($r$ being radius of the nanostructure, and $q$ as wave vector). The dielectric constant of the air medium is taken as 1. We use 10.4 and 5.8 corresponding to the values of $\varepsilon_0$ and $\varepsilon_\infty$, respectively, for GaN.[26] SO phonons associated with $A_1$(TO) at 534 cm$^{-1}$ and $E_1$(TO) at 561 cm$^{-1}$ are calculated for GaN. Fig. 6 shows the $\omega_{SO}$ for GaN as a function of $qr$ following Eq. 1. It is interesting to note that the calculated values of SO phonons pertaining to the $E_1$ character of GaN are very close to the new peak observed around 690 cm$^{-1}$ (Fig. 4c) at $qr = 0.99$ in our study with 632.8 nm excitation (Fig. 6). Similarly, the broad peak in the range of 665–699 cm$^{-1}$ (inset in Fig. 4d) may be assigned to the combination of SO modes. At $qr = 1.18$, we get 665 cm$^{-1}$ and 699 cm$^{-1}$ as a SO phonons pertaining to the $A_1$ and $E_1$ characters of GaN, respectively (Fig. 6). However, at $qr = 1.18$, we also get a peak around 680 cm$^{-1}$, which can be related to the $E_1$ character of In$_{0.15}$Ga$_{0.85}$N considering $E_1$(TO) around 545 cm$^{-1}$ which (inset in Figs. 4c, and 4d).[15] In the absence of available values of dielectric constants for specific composition of In$_{0.15}$Ga$_{0.85}$N, we assume similar $qr$ dependence of the SO modes of InGaN. The calculated value is around 672 cm$^{-1}$, which is close to the observed value of 680 cm$^{-1}$. The small discrepancy in the calculation can perhaps be removed with exact determination of dielectric constants for $In_{0.15}Ga_{0.85}N$ phase. Thus, the origin of the broad new peak in the range of 665–699 cm$^{-1}$ can be assigned to the SO modes corresponding to 1-D GaN and $In_{0.15}Ga_{0.85}N$. We must restate here that the broad peak in InGaN around 700 cm$^{-1}$ is identified as the specific mode 'S' without any physical implications.[15,24,25] We must also mention here that report of the SO mode at 716 cm$^{-1}$ for GaN columnar structures of transverse width 50 nm using similar dielectric constants and $\lambda_{Ex}$ = 488 nm ($qr$=0.64)[11] is much higher than our estimate (636 cm$^{-1}$ as $A_1$ and 665 cm$^{-1}$ as $E_1$ characters from Fig. 6). A broad size distribution, specific to the reported anodization technique, in the columnar structures on the GaN thin film, and defects present in the system might have suppressed these peaks. The SO phonon dispersion largely depends both on the dielectric function of the semiconductor and that of the surrounding medium in contact. Taking this fact into account, typical Raman spectra for nanopillars in a higher dielectric medium ($\varepsilon_m$ = 2.56 for aniline) are recorded and compared with those in air for the same excitation of a 532 nm laser line (Fig. 7). One can see that the peak positions of the $A_1(LO)_{GaN}$ and $A_1(LO)_{InGaN}$ modes do no change, while a clear red-shift of ~15 cm$^{-1}$ for SO modes for GaN is observed for the spectrum collected from the sample surrounded with a higher value of dielectric constant than that of air ($\varepsilon_m$ =1). The intensities of the observed SO phonons in the present study are reasonably high with respect to other phonons. The magnitude of surface roughness determines the SO phonon intensities owing to the breakdown of the translational symmetry of the surface potential.[7,12,27] The wavelength $\lambda = 2\pi/q$ corresponding to the surface potential (perturbation) can be estimated as ~ 250–300 nm for the nanopillars (using the dispersion relations in Fig. 6 with various excitations). Surface roughness with a modulation of ~25 nm is observed (Fig. 5) in these nanostructures. Modulation length with any integral multiple equivalent to the $\lambda$ (10x25 nm=250 nm or 12x25 nm=300 nm) will be sufficient to initiate breakdown of translational symmetry for the contribution of surface potential towards SO phonons and makes the intensity of the surface modes comparable to that of the other phonons.

**CONCLUSION**

FIB direct milling of InGaN/GaN MQW epi-layers was performed with the site-controlled realization of high-aspect-ratio nanopillars, which were size-controlled by wet-etching. The retained QW discs inside the pillars emit a sharp luminescence peak, with the peak intensity greatly enhanced upon subsequent KOH treatment. Enhanced luminescence intensity and corresponding phonon properties supported the results at the molecular level, and the SO phonon modes for the ternary InGaN phase are identified for the first time. The results indicate that the removal of the surface amorphous layer is crucial for FIB micro/nano engineering on III-nitride-based materials, and thus provide an effective method for size-controlled nanorod array fabrication, paving the way for FIB prototyping of III-nitride optical devices.

**Acknowledgements**

The authors gratefully acknowledge the financial support of the National Science Council of Taiwan, under grants NSC96-2815-C-006-021-E, NSC95-2221-E-006-079-MY3 and NSC95-2120-M-006-010. The authors would also like to thank the NSC Core Facilities Laboratory for Nano-Science and Nano-Technology in the Kaohsiung-Pingtung Area for equipment access and technical support. One of the authors (SD) acknowledges Y. Tzeng of the Institute of Innovations and Advanced Studies (IIAS), NCKU for his encouragement during this work.

**References**

1. Dai D, Zhang B, Lin J Y, Jiang H X. *J. Appl. Phys*. 2001; **89**: 4951.

2. Choi H W, Dawson M D, Edwards P R, Martin R W. *Appl. Phys. Lett.*, 2003; **83**: 4483.

3. Hsueh T H, Huang H W, Lai F I, Sheu J K, Chang Y H, Kuo H C, Wang S C. *Nanotechnology* 2005; **16**: 448.

4. Kim H M, Cho Y-H, Lee H, Kim S I, Ryu S R, Kim D Y, Kang T W, Chung K S. *Nano Lett.* 2004; **4**: 1059.

5. Chiu C H, Lu T C, Huang H W, Lai C F, Kao C C., Chu J T, Yu C C, Kuo H C, Wang S C, Lin C F, Hsueh T H. *Nanotechnology* 2007; **18**: 445201.

6. Tseng A A. *J. Micromech. Microeng.* 2004; **14**: R15.

7. Adu K W, Xiong Q, Gutierrez H R, Chen G, Eklund P C. *Appl. Phys. A* 2006; **85**: 287.

8. Ruppin R, Englman R. *Rep. Prog. Phys.* 1970; **33**: 149.

9. Sernelius B E., *Surface Modes in Physics*, Wiley-VCH, New York, 2001.

10. Hsiao C L, Tu L W, Chi T W, Chen M, Young T F, Chia C T, Chang Y M. *Appl. Phys. Lett.* 2007; **90**: 043102.

11. Tiginyanu I M, Sarua A, Irmer G, Monecke J, Hubbard S M, Pavlidis D, Valiaev V. *Phys. Rev. B* 2001; **64**: 233317.

12. Sahoo S, Hu M S, Hsu C W, Chen L C, Chen K H, Arora A K, Dhara S. *Appl. Phys. Lett.* 2008; **93**: 233116.

13. Wu S E, Liu C P, Hsueh T H, Chung H C, Wang C C, Wang C Y. *Nanotechnology* 2007; **18**: 445301.

14. Azuhata T, Sota T, Suzuki K, Nakamura S. *J. Phys.*: *Condens. Matter* 1995; **7**:

L129.

15. Kontos A G, Raptis Y S, Pelekanos N T, Georgakilas A, Bellet-Amalric E, Jalabert D. *Phys. Rev.* B 2005; **72**: 155336.

16. Porto S P S, Krishnan R S. *J. Chem. Phys.* 1967; **47**: 1009.

17. Zi J, Wan X, Wei G, Zhang K, Xie X. *J. Phys. Condens. Matter* 1996; **8**: 6323.

18. Chen C C, Yeh C C, Chen C H, Yu M Y, Liu H L, Wu J J, Chen K H, Chen L C, Peng J Y, Chen Y F. *J. Am. Chem. Soc.* 2001; **123**: 2791.

19. Siegle H, Kaczmarczyk G, Filippidis L, Litvinchuk L, Hoffmann A, Thornsen C. *Phys. Rev. B* 1997; **55**: 7000.

20. Correia M R, Pereira S, Pereira E, Frandon J, Alves E. *Appl. Phys. Lett.* 2003; **83**: 4761.

21. Lazić S, Moreno M, Calleja J M, Trampert A, Ploog K H, Naranjo F B, Fernandez S, Calleja E. *Appl. Phys. Lett.* 2005; **86**: 061905.

22. Grille H, Schnittler C, Bechstedt F. *Phys. Rev. B* 2000; **61**: 6091.

23. Alexson D, Bergman L, Nemanish R J, Dutta M, Stroscio M A, Parker C A, Bedair S M, El-Masry N A, Adar F. *J. Appl. Phys.* 2001; **89**: 798.

24. Correia M R, Pereira S, Pereira E, Frandon J, Watson I M, Liu C, Alves E, Sequeira A D, Franco N. *Appl. Phys. Lett.* 2004 ; **85**: 2235.

25. Tabata A, Teles L K, Scolfaro L M R, Leite J R, Kharchenko A, Frey T, As D J, Schikora D, Lischka K, Furthmuller J, Bechstedt F. *Appl. Phys. Lett.* 2002; **80**: 769.

26. Madelung O. *Semiconductors: Data Handbook*, 3rd ed., Springer, Berlin, 2004; section **2.9**.

27. Gupta R, Xiong Q, Mahan G D, Eklund P C. *Nano Lett.* 2003; **3**: 1745.

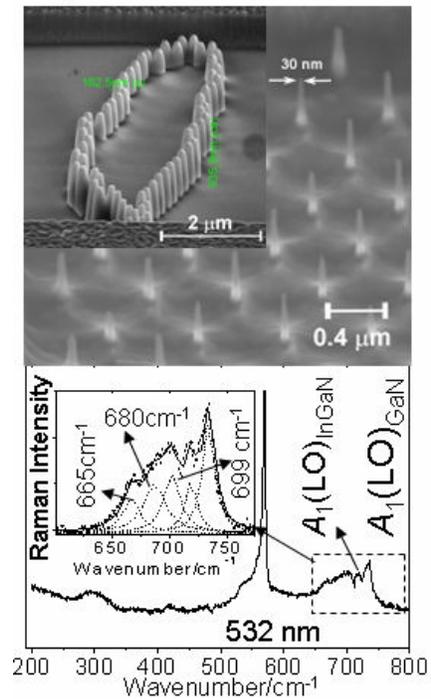

**Figure Captions**

**Figure TOC.** Formation of InGaN/GaN multi-quantum well (MQW) nanopillar arrays using focused ion beam irradiation and subsequent chemical etching. Crystalline MQW nanopillars with diameter, as low as 30 nm and high aspect ratio of 16:1 are reported along with a three-fold enhancement in emission intensity per unit active area. Raman scattering studies ware used to characterize the newly created surfaces. Unknown peaks in GaN and InGaN nanostructures are identified for surface optical modes with proper assignments of wave vectors using multiple excitations.

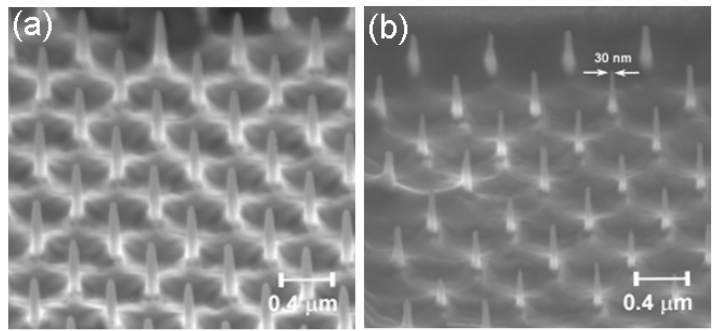

**Figure 1.** a) SEM tilted-view images showing FIB milled nanopillars after KOH treatment. b) A smaller pillar size of 30 nm achieved with a longer KOH treatment time than that used in the previous.

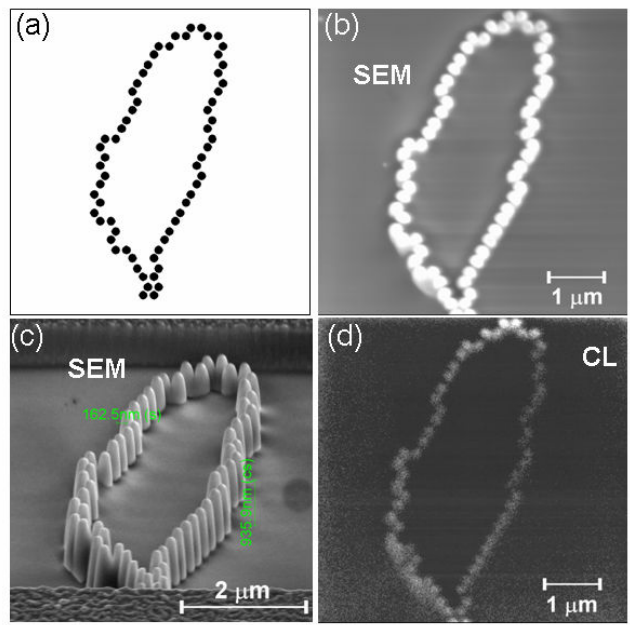

**Figure 2.** a) Room temperature CL spectra acquired from the pristine, as-irradiated and chemically etched samples. The emission intensities from the arrayed pillars and the pristine sample were normalized by the effective quantum well area. b) A typical SEM and corresponding CL image at the specific wavelength of 420 nm of the nanopillar array.

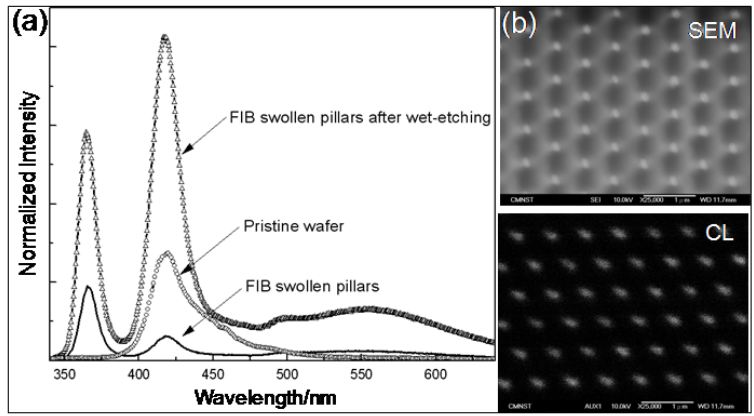

**Figure 4.** Polarized Raman scattering studies of (a) a pristine MQW epi-layer; b) as-milled pillars. Spectra for wet chemical (KOH) etched nanopillars with the excitations of (c) 632.8 nm with inset showing the detailed spectrum in the 500-600 cm$^{-1}$ range and (d) 532 nm laser lines with inset showing the Lorentzian fits for multiple peaks in the 625-775 cm$^{-1}$ range.

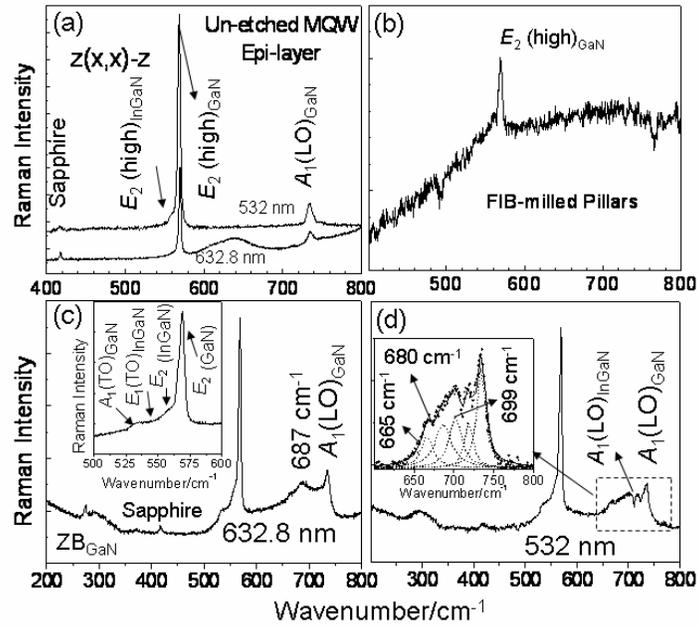

**Figure 3.** (Color online) a) Diagram of the bitmap file used for FIB patterning. Pixels in the black regions were not exposed to the FIB. b) SEM top-view of a contour map of Taiwan grown with the nanopillar structures by FIB milling and subsequent reactive ion etching. c) SEM lateral-views showing the nanopillars constituting the map. d) CL image of the top-view of the light emitting Taiwan.

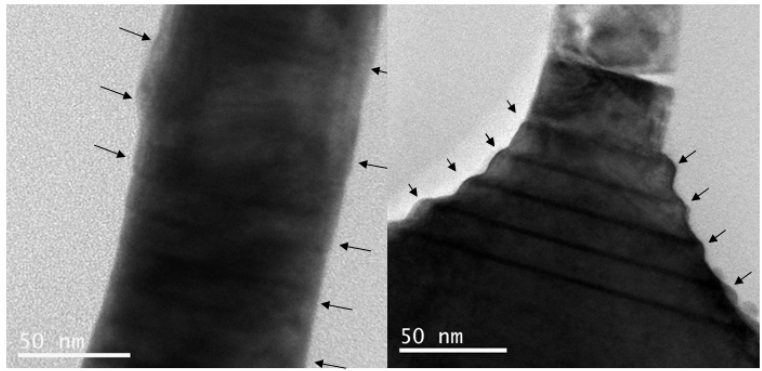

**Figure 5.** Surface roughness in typical MQW nanopillars. Arrows show the modulation in the surface structure at two different regions.

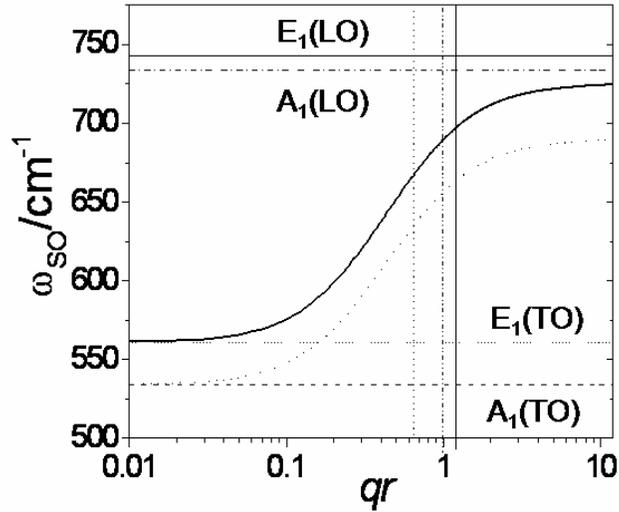

**Figure 6.** Calculated SO phonon modes of GaN as a function of $qr$, full curve: SO($E_1$), dashed curve: SO($A_1$), horizontal full and dashed lines are the TO and LO wavenumbers of $E_1$ and $A_1$ modes, respectively. Vertical lines are marked for $qr$=0.64 (dotted), $qr$=0.99 (dash-dotted) and $qr$=1.18 (full).

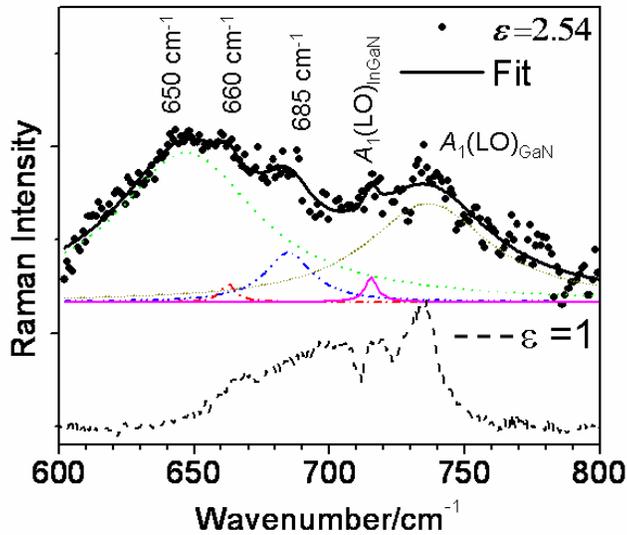

**Figure 7.** (Color online) Raman spectra for wet etched nanopillars immersed in aniline ($\varepsilon$=2.54) and in air ($\varepsilon$=1) with the excitation of the 532 nm laser line. Lorentzian fits for multiple peaks are also shown for the spectrum recorded with anailine. Individual peaks are also shown with broken lines.

18